\documentclass{aa}
\usepackage[varg]{txfonts}
\usepackage{multirow}
\usepackage{twoopt}
\usepackage{amsmath}
\usepackage{mathtools}
\usepackage{xr}
\usepackage{color}
\usepackage[normalem]{ulem}
\usepackage[breaklinks=true]{hyperref} 
\bibpunct{(}{)}{;}{a}{}{,}             
\makeatletter
  \newcommandtwoopt{\citeads}[3][][]{\href{http://adsabs.harvard.edu/abs/#3}%
    {\def\hyper@linkstart##1##2{}%
     \let\hyper@linkend\@empty\citealp[#1][#2]{#3}}}
  \newcommandtwoopt{\citepads}[3][][]{\href{http://adsabs.harvard.edu/abs/#3}%
    {\def\hyper@linkstart##1##2{}%
     \let\hyper@linkend\@empty\citep[#1][#2]{#3}}}
  \newcommandtwoopt{\citetads}[3][][]{\href{http://adsabs.harvard.edu/abs/#3}%
    {\def\hyper@linkstart##1##2{}%
     \let\hyper@linkend\@empty\citet[#1][#2]{#3}}}
  \newcommandtwoopt{\citeyearads}[3][][]%
    {\href{http://adsabs.harvard.edu/abs/#3}
    {\def\hyper@linkstart##1##2{}%
     \let\hyper@linkend\@empty\citeyear[#1][#2]{#3}}}
\makeatother

\begin{document}
\title{Search for grain growth towards the center of L1544 \thanks{Based on observations carried out with the IRAM 30m Telescope. IRAM is supported by INSU/CNRS (France),MPG (Germany) and IGN (Spain)} }

\author{A.~Chac\'on-Tanarro \inst{1}
\and P.~Caselli \inst{1}
\and L.~Bizzocchi \inst{1}
\and J.~E.~Pineda \inst{1}
\and J. Harju \inst{1,2}
\and M. Spaans \inst{3}
\and F.-X. D\'esert \inst{4}
}
\institute{Max-Planck-Instit\"{u}t f\"{u}r extraterrestrische Physik, Giessenbachstrasse 1, 85748 Garching, Germany
\and Department of Physics, P.O. Box 64, 00014 University of Helsinki, Finland
\and Kapteyn Astronomical Institute, University of Groningen, P.O. Box 800, 9700 AV Groningen, The Netherlands
\and Univ. Grenoble Alpes, CNRS, IPAG, Grenoble F-37000, France
}
\date{Received - / Accepted -}

\abstract{
In dense and cold molecular clouds dust grains are surrounded by thick icy mantles. It is however not clear if dust growth and coagulation take place before the switch-on of a protostar. This is an important issue, as the presence of large grains may affect the chemical structure of dense cloud cores, including the dynamically important ionization fraction, and the future evolution of solids in protoplanetary disks. To study this further, we focus on L1544, one of the most centrally concentrated pre-stellar cores on the verge of star formation, and with a well-known physical structure. We observed L1544 at 1.2 and 2 mm using NIKA, a new receiver at the IRAM 30 m telescope, and we used data from the Herschel Space Observatory archive. We find no evidence of grain growth towards the center of L1544 at the available angular resolution. Therefore, we conclude that single dish observations do not allow us to investigate grain growth toward the pre-stellar core L1544 and high sensitivity interferometer observations are needed. We predict that dust grains can grow to 200 $\mu$m in size toward the central $\sim$300 au of L1544. This will imply a dust opacity change by a factor of $\sim$2.5 at 1.2 mm, which can be detected using the Atacama Large Millimeter and submillimeter Array (ALMA) at different wavelengths and with an angular resolution of 2\arcsec. }


\maketitle

\section{Introduction}

Pre-stellar cores are self-gravitating starless dense cores with clear signs of contraction motions and chemical evolution \citepads{2005ApJ...619..379C}. They are formed within molecular clouds, due to the influence of gravity, magnetic fields and turbulence. They are considered to be on the verge of star formation, and therefore represent the initial conditions in the process of star formation \citepads{2007ARA&A..45..339B, 2012A&ARv..20...56C}. These systems are characterized by high densities ($n_{H_2}>10^5$\,cm$^{-3}$) and low temperatures ($T<10$\,K) towards their central regions \citepads{2007A&A...470..221C}. 

In dense and cold molecular clouds dust grains are surrounded by thick icy mantles \citep[e.g.][and references therein]{2015ARA&A..53..541B}. It is however not clear if dust growth and coagulation take place before a protostar is born. This is an important issue, as dust coagulation may affect the formation and evolution of protoplanetary disks, forming from molecular clouds \citepads{2016MNRAS.460.2050Z}. \citetads{2013MNRAS.428.1606F} find a strong correlation between the visual extinction and the slope of the extinction law toward the Perseus molecular cloud. This can be interpreted as grain growth, but it is not clear if grain coagulation is needed or if the growth of icy mantle can explain the observed correlation. Large grains are detected also in young protoplanetary disks, again suggesting that grain growth may already be at work in the earlier dense core phases \citepads{2014prpl.conf..339T}. There is evidence of large (micrometer-sized) grains through the observed extended emission at 3.6 $\mu$m in dense cloud cores \citep[the so-called coreshine effect]{2010Sci...329.1622P}, but this interpretation has been questioned by \citetads{2013A&A...558A..62J}, who suggested that amorphous hydrocarbon material could actually produce the observed coreshine without the need of large grains.  Moreover, \citetads{2014A&A...568L...3A} found a similar threshold for the coreshine and water ice, with the scattering efficiency at 3.6 $\mu$m increasing with the increase of the water-ice abundance, suggesting that water ice mantle growth in dense clouds may be at least partially responsible for the coreshine effect. 

The study of the dust emission at long wavelengths, sensitive to the larger grains, is therefore important to gain a better understanding on the grain size distribution on the early phases of star formation. This emission depends on dust grains properties such as structure, size and composition. Here we will focus on the dust opacity and the spectral index, two parameters that depend directly on the grain size distribution. 

The opacity, $\kappa_{\nu}$, is a measurement of the dust absorption cross sections weighted by the mass of the gas and dust. As explained by \citetads{1994A&A...288..929K}, when dust grains are small compared to the wavelength, the so-called Rayleigh limit, the mass absorption coefficient does not depend on the grain size, but only on the mass, when the grain size is much larger than the wavelength, the absorption coefficient depends inversely on the grain size, and when $ a\sim\lambda$ ($a$ being the grain radius), the mass absorption coefficient can increase up to 10 times its value, because at these wavelengths dust grains are better radiators \citepads{1994A&A...288..929K}. When dust grains are coated by ices, their cross section increases and, consequently, so does $\kappa_{\nu}$. This same trend is seen with fluffy dust grains \citepads{1994A&A...288..929K, 1994A&A...291..943O}. Therefore, variations on the value of $\kappa_{\nu}$ along the spectrum can be a strong indicator of grain growth. 

The dust opacity can be approximated by a power-law at millimeter wavelengths, $\kappa_{\nu}\sim\nu ^{\beta}$, where $\beta$ is the emissivity spectral index \citepads{1983QJRAS..24..267H}. For $\beta$, a more complex analysis is needed. Typical values found in the interstellar medium (ISM) at far-infrared and sub-millimeter wavelengths lie in the range 1.5-2. In presence of large grains, which increase the dust opacity at longer wavelengths, $\beta$ can decrease to values close to or below 1 \citepads{1994A&A...291..943O, 2006ApJ...636.1114D}. When the minimum size of the distribution is lower than 1 $\mu$m, $\beta$ is only affected by the largest grains, and not by the smaller grains \citepads{1993Icar..106...20M, 2006ApJ...636.1114D}. 

Laboratory measurements of the opacity and the spectral index at millimeter wavelengths found a dependence of the mass absorption with temperature for different grain compositions \citepads{1996ApJ...462.1026A, 1998ApJ...496.1058M,2005ApJ...633..272B,2011A&A...535A.124C,2013lcdu.confE..44D}.  \citetads{1996ApJ...462.1026A} found two different behaviors depending on the temperature range: for very low temperatures (1.2 - 20 K) the millimeter opacity decreases with increasing temperature, while for temperatures between 20 and 30 K, it increases or is constant with temperature. This trend turns out to be the opposite for $\beta$. However, the measured changes are not significant (within 20\%) at mm-wavelengths, when the temperature varies between 6 and 10 K, the temperature range relevant for the central regions of pre-stellar cores. \citetads{2005ApJ...633..272B} and \citetads{2011A&A...535A.124C} found similar results for amorphous material, as they observed an increase in the spectral index (decrease in the opacity) while decreasing the temperature (for $T >10$ K). However, when they let the material crystallize, no temperature dependence was detected. Additionally, they also reported a frequency-dependence on the opacity and the spectral index. Another interesting result was found by \citetads{2013lcdu.confE..44D}, who showed two extreme and completely different results depending on the material analyzed for temperatures ranging from 10 K to 300 K. For wavelengths longer than 500 $\mu$m, the spectral index value could take values larger than 2.5 and smaller than 1.5 depending on the composition of the sample, which in their case depends on the oxidation state of the iron. This means that the increase on the opacity in the studied temperature range is more important at longer wavelengths. What \citetads{2013lcdu.confE..44D} conclude is that the emission from dust cannot be described by only one power-law spectrum, but it should be characterized by different spectral indexes at different wavelengths. 

There are several astronomical studies constraining the value of $\beta$ in protoplanetary disks, where it is known that dust coagulation is taking place to give birth to future planets. \citetads{2004ASPC..323..279N} found that, despite the composition dependence and grain distribution shape, dust grains of 1.2 mm of size lead to $\beta$ values lower than 1. For earlier stages in the process of star formation,  \citetads{2014MNRAS.444.2303S} found low values for the spectral index towards OMC 2/3 with a low anticorrelation between $\beta$ and temperature, which may indicate the presence of millimeter size grains up to centimeter size. However, this was subject of further study by \citetads{2016A&A...588A..30S}, who found higher values of the spectral index and suggested that the observations from \citetads{2014MNRAS.444.2303S} were contaminated or deviated from a single power-law. Studying extinction maps, \citetads{2015A&A...580A.114F} found that, while a single spectral index can reproduce their observational data, the opacity increases towards the center of the starless core FeSt 1-457, possibly indicating grain growth. At larger scales, using Planck and Herschel results, \citetads{2015A&A...584A..94J, 2015A&A...584A..93J} observed that the opacity increases with density and that there is an anticorrelation between the spectral index and the temperature. 

Obtaining a value of the spectral index at early stages of star formation is a difficult task due to the known degeneracy between the spectral index and the temperature, which appears when the spectral energy distribution of the dust emission is fitted using least squares to obtain the temperature, density, opacity and spectral index of the observed object \citepads{2009ApJ...696..676S, 2009ApJ...696.2234S}. Moreover, it has also been proved that uncertainties in the measured fluxes, and  an incorrect assumption of isothermality and the noise itself can mimic the observed anticorrelation between the temperature and the spectral index of the dust \citepads{2009ApJ...696..676S, 2009ApJ...696.2234S}. 

Therefore, taking into account these studies and previous results on the spectral index, it is not clear what to expect from observations towards the dense and cold pre-stellar cores. Nevertheless, any variation of the $\kappa_{\nu}$ and/or $\beta$ across a cloud core, from the outskirts to the center, would indicate grain growth. \citetads{2015A&A...579A..15K} show that dust evolution in dense clouds produces significant variations (factors of a few) of the opacity, while the spectral index only changes by less than 30\%, so that $\kappa_{\nu}$ variations should be easier to measure. Moreover, if the physical structure of the cloud is known, one could directly measure the variation of the opacity across a core at millimeter wavelengths, while for the spectral index a multi-wavelength study is needed. 

In this work we focus on L1544, a well known pre-stellar core in the Taurus Molecular Cloud at a distance of 140 pc. The zone within the central 1000 au is still unexplored, but we know that the temperature is dropping down to 7 K toward the central 2000 au \citepads{2007A&A...470..221C} and it shows clear signs of contraction \citepads{2012ApJ...759L..37C}. Detailed modeling of L1544 \citepads{2010MNRAS.402.1625K} found that an increase in the dust opacity is needed to reproduce the drop in the measured temperature toward the central 2000 au. This could be an indication of fluffy grains in the core center \citepads{1994A&A...291..943O, 2009A&A...502..845O}, where CO is heavily frozen \citepads{1999ApJ...523L.165C} and volume densities become larger than $10^{6}$ cm$^{-3}$. The presence of fluffy grains can only be verified by multiwavelength millimeter observations. The well-known physical structure of L1544, with its high volume densities and centrally concentrated structure, makes this object the ideal target to study possible variation in the opacity, using the continuum emission at 1.2 mm and 2 mm from the IRAM 30 m telescope, with an angular resolution of 12.5\arcsec and 18.5\arcsec, respectively (at 140 pc, this corresponds to 1800 and 2600 au). This is the first study of opacity variation and grain growth across a pre-stellar core with such resolution and at such long wavelengths using a single dish telescope. 

The paper is organized as follows. In Section \ref{Observations} we describe the millimeter data for L1544 obtained with NIKA and far-IR data obtained with SPIRE. In Section \ref{dust-properties} we describe the results for the maps of the spectral index and the opacity assuming constant temperature and density across the cloud, as done in previous studies. This is compared later with the same results assuming variations on both the temperature and the density along the line of sight, making use of the profiles presented by \citetads{2015MNRAS.446.3731K}. We end this section presenting the results for the opacity and spectral index from the combination of NIKA and Herschel/SPIRE data, and a comparison between the modeled emission and the observations. In Section \ref{model_sec} we compare our observational findings with a simple grain growth evolution model. In Section \ref{conclusions} we summarize our findings.

\section{Observations} \label{Observations}

\subsection{NIKA}
The observations were carried out using the IRAM 30 m telescope, located at Pico Veleta (Spain), during the spring of 2014, using the New IRAM KID Array (NIKA) \citepads{2014A&A...569A...9C, 2013A&A...551L..12C}. The project number is 151-13. A region of 3\degr $\times$ 3\degr was mapped using the Lissajous pattern to observe the pre-stellar core L1544, $\alpha$(J2000) = 05$^{\mathrm{h}}$
04$^{\mathrm{m}}$17.21$^{\mathrm{s}}$ and $\delta$(J2000) = $+$25\degr 10\arcmin 42.8\arcsec \citep[dust peak from]{1999MNRAS.305..143W}, at 1.2 and 2 mm. The main beam widths are 12.5\arcsec at 1.2 mm and 18.5\arcsec at 2 mm. At a distance of 140 pc, this corresponds to a resolution of 1800 au and 2600 au, respectively. The KID array has a field-of-view is 1.8\arcmin at 1.2 mm and 2.0\arcmin at 2 mm. 

\begin{figure} 
\resizebox{\hsize}{!}{\includegraphics{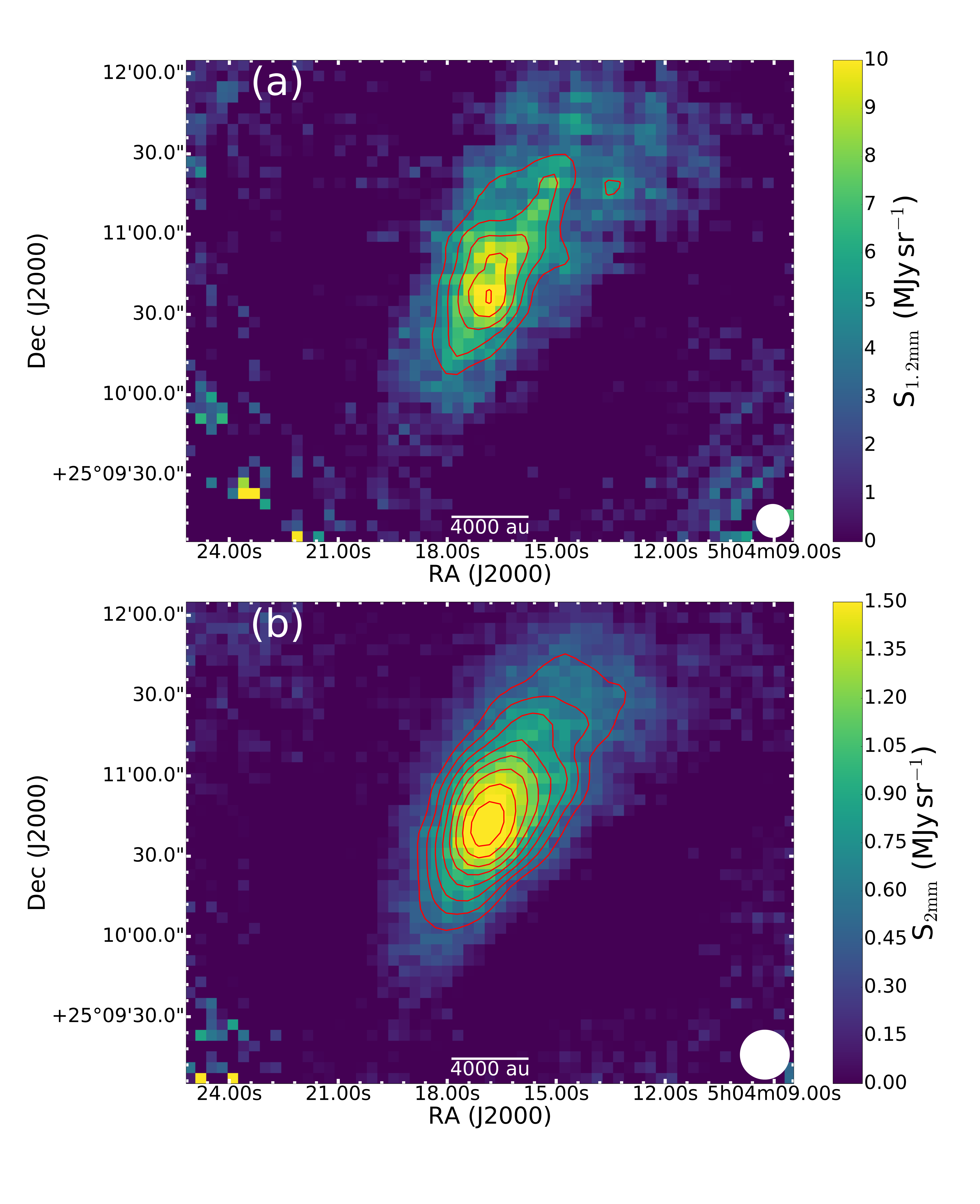}} 
\caption{Maps of the continuum emission of L1544 at (a) 1.2 mm and (b) 2 mm. The angular resolution is 12.5\arcsec at 1.2 mm and 18.5\arcsec at 2 mm (HPBW shown by the white circle in the bottom right corner) and the contours are indicating the different $\sigma$-levels of the emission above 3 $\sigma$ (3 $\sigma$, 4 $\sigma$, 5 $\sigma$,  etc., where $\sigma_{1.2mm} = 1.4$ MJy/sr and $\sigma_{2mm} = 0.15$ MJy/sr). }
\label{Fluxes}
\end{figure}

The data were reduced by the NIKA team. The original maps have been corrected by the main beam to full beam ratio, which is 1.56 at 1.2 mm and 1.35 at 2 mm, with an uncertainty of $\sim$10\% on these factors. The noise level of the map at 1.2 mm is 0.007 Jy/12.5\arcsec-beam and that of the map at 2 mm is 0.002 Jy/18.5\arcsec-beam. In Fig. \ref{Fluxes}, both images have been converted into MJy/sr. In Sections 3.2.1 and 3.2.2, the two maps have been convolved to a common resolution of 19\arcsec, so the final noise levels are 0.9 and 0.15 MJy/sr at 1.2 and 2 mm respectively. The absolute uncertainties on the surface brightnesses are $\sim$15\%. 

The maps show two lobes of negative surface brightness: one to the north-east and another one to the south-west, due to filtering of extended emission. As the field-of-view of the camera differs by $\sim$10\% between the two NIKA bands, we expect this filtering to be the same for both bands, and therefore, not to affect our analysis done in Section 3.2, where both bands are treated simultaneously. To quantify this statement, a study of the power spectra of both bands is performed in Appendix A. Moreover, \citetads{2015A&A...576A..12A} used simulations to estimate at which scales NIKA filters the emission. They found that NIKA filtering starts to be important at scales larger than the field-of-view and that it is the same for the two bands. This is also mentioned in \citetads{2017A&A...598A.115A}. Thus, here filtering effects are also assumed to be within the errors for Section 3.3, where results are shown for each band separately, and only for the region above 3 $\sigma$ detection, which is $\sim$1\arcmin  in size, and therefore, smaller than the field-of-view.  

\subsection{Herschel}

The Herschel/SPIRE data at 250, 350, and 500 $\mu$m used here were already presented by \citetads{2016arXiv160703242S}.  They are part of the Level 3 mosaic maps of the Taurus molecular cloud complex observed in the SPIRE/PACS parallel mode in the course of the Herschel Gould Belt Survey \citepads{2010A&A...518L.102A}. The data downloaded from the Herschel Science Archive (HSA) are reduced using the Standard Product Generation (SPG) software version 14.2.1. These data are calibrated assuming that the source is infinitely extended, and the surface brightness zero points have been corrected using absolute levels from Planck maps. The SPIRE flux densities are assumed to be accurate to 10\% \citepads{2016A&A...588A.107B}.  The resolution of Herschel at 500 $\mu$m is $\sim$38.5\arcsec. When combining the SPIRE maps with the NIKA data, we took into account that SPIRE is sensitive to the extended emission, while NIKA filters it out. To compensate for this difference, we subtracted from each SPIRE map a value which corresponds to the average surface brightness just outside the region seen by NIKA.  The region where these averages are calculated have been determined from the NIKA 1.2 mm map; this is a $\sim$35 arcsec wide ring outside the 1 $\sigma$ contour. After this subtraction, the peak emission and the errors at 250, 350, and 500 $\mu$m are 153$\pm$12, 120$\pm$12 and 62$\pm$9 MJy/sr, respectively. The dependence of our results on the width of the ring was checked, and found not significant.

\section{Dust properties} \label{dust-properties}

\subsection{Theoretical background}

Dust grains emit as modified black bodies:

\begin{equation}
S_{\nu} = \Omega B_{\nu}(T_{d}) \kappa_{\nu} \mu_{H_2} m_{H}N(H_{2}),  
\label{eq1}
\end{equation}
\\
where $B_{\nu}(T_{d})$ is the black body function at a temperature $T_{d}$:
\begin{equation}
B_{\nu}(T_{d}) = \frac{2h\nu^3}{c^2} \frac{1}{\exp \left(\frac{h\nu}{kT_{d}}\right)-1},
\end{equation}
\\
$S_{\nu}$ is the flux density, $\Omega$ is the solid angle subtended by the beam, $\kappa_{\nu}$ is the dust opacity at frequency $\nu$ (the absorption cross-section for radiation per unit mass of gas), $\mu_{H_2}=2.8$ is the molecular weight per hydrogen molecule \citepads{2008A&A...487..993K}, $m_{H}$ is the mass of the hydrogen atom and $N_{H_{2}}$ is the molecular hydrogen column density. The dust opacity can be approximated by a power-law at mm-wavelengths \citepads{1983QJRAS..24..267H}:

\begin{equation}
\kappa_{\nu} = \kappa_{\nu_0}\left( \frac{\nu}{\nu_{0}}\right)^{\beta},
\end{equation}
\\
where $\kappa_{\nu_0} $ is the opacity at a reference frequency $\nu_0$ and $\beta$ is the spectral index. We assume a dust-to-gas ratio of 0.01 \footnote{Although different dust-to-gas mass ratios have been measured toward different directions in our Galaxy \citepads{2011piim.book.....D}, we do not expect significant changes within the same cloud.} in Eq. \eqref{eq1}, in which there are several unknown parameters: the density, the temperature, the opacity and the spectral index. 

The above equations are typically used to derive dust temperatures, densities and masses, assuming isothermal clouds and using fixed values for the dust opacity and the spectral index \citep[e.g.]{2005ApJ...624..254S}. The value of the dust opacity is normally taken from Table 1 of \citetads{1994A&A...291..943O}, and the spectral index is mostly assumed to be within the range of 1.5-2 for pre-stellar cores. However, these approximations and assumptions can give wrong temperatures, densities and masses or false anticorrelations between $\beta$ and the dust temperature \citepads{2009ApJ...696..676S, 2009ApJ...696.2234S}.

Here we first follow the calculation of the spectral index done by \citetads{2005ApJ...624..254S} assuming a constant temperature of 10\,K for the core. We then do the same calculation but using the temperature and density profiles derived by \citetads{2015MNRAS.446.3731K} and compare them to the constant temperature case. Finally, we consider a constant value of the spectral index and the opacity across the core to see if the observed millimeter emission can be reproduced by modeling. 

\subsection{Spectral index and opacity of the dust using NIKA}
\subsubsection{Assuming constant temperature}

For isothermal clouds, the spectral index value can be derived using the ratio of the surface brightnesses at 1.2 and 2 mm:

\begin{equation}
\beta = \frac{\log \frac{S_{1.2mm}}{S_{2mm}} - \log \frac{B_{1.2mm}(T_d)}{B_{2mm}(T_d)}}{\log \frac{\nu_{1.2mm}}{\nu_{2mm}}} .
\label{ratio}
\end{equation}

Considering a constant $\kappa _{250 \mu m}$=0.1 cm$^2$g$^{-1}$ value \citepads{1983QJRAS..24..267H}, we can also derive a map for the opacity at 1.2 mm:
\begin{equation}
\kappa_{1.2 mm} = \kappa_{250\mu m}\left( \frac{\nu_{1.2 mm}}{\nu_{250 \mu m}}\right)^{\beta} .
\end{equation}

The values have been only derived for surface brightnesses above the 3 $\sigma$ level of the maps at 1.2 and 2 mm. 

\begin{figure} 
\resizebox{\hsize}{!}{\includegraphics{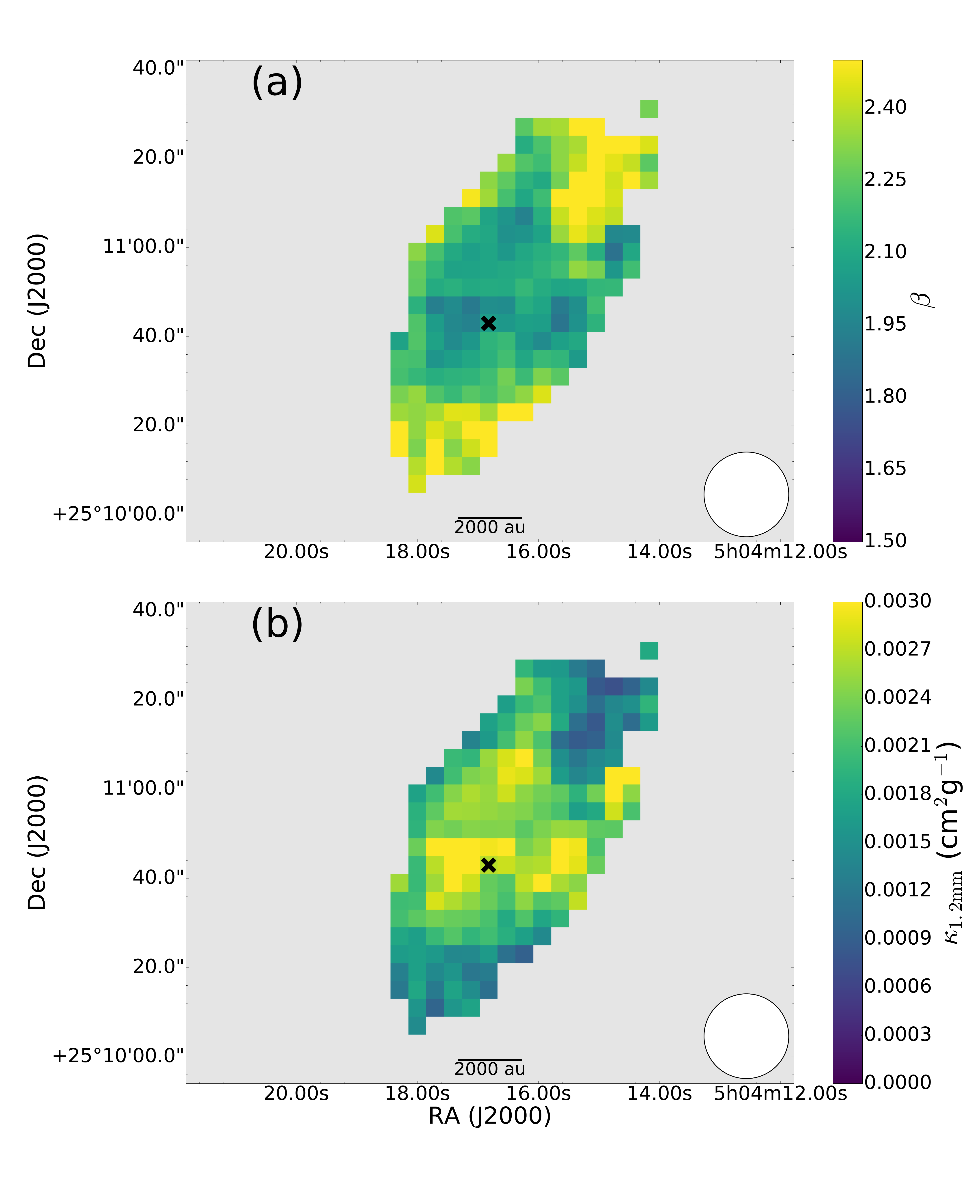}} 
\caption{Maps of the (a) spectral index and (b) the opacity at 1.2 mm assuming a constant core temperature of 10 K and a constant $\kappa _{250 \mu m}$=0.1 cm$^2$g$^{-1}$ value \citetads{1983QJRAS..24..267H}. The 1.2 mm map was convolved to the 2 mm band resolution, which is 19\arcsec (HPBW shown by the white circle in the bottom right corner). In both cases, the crosses mark the peak of the emission corresponding to the final 1.2 mm map convolved to 19\arcsec.}
\label{kappa-beta-10k}
\end{figure}

The resulting $\beta$ and $\kappa_{1.2mm}$ maps are shown in Fig. \ref{kappa-beta-10k}. There is a slight decrease of the spectral index towards the central regions, although its value, $\sim$2, is still consistent with interstellar medium (ISM) dust, not affected by coagulation.  This observed spatial gradient, which shows variations of $\sim$30\%, appreciable with an errorbar of $\sim$25\% in the spectral index value, is reflecting the variation of the $S_{1.2mm}/S_{2mm}$ ratio across the core, as the other quantities in Eq. \eqref{ratio} are constant. Therefore, an increase of the temperature in the outskirts would necessarily mimic an increase of $\beta$. 

The opacity shows a significant increase towards the center of a factor of 10, with errorbars of $\sim$0.001 cm$^2$g$^{-1}$ in the determination of the opacity values. This could indicate grain growth. However, this result is also affected by the underlying assumption that the temperature is constant across the core, while this is not the case for L1544 (see Section 3.2.2). Moreover, $\kappa_{250 \mu m}$ may vary as well across the cloud, so the assumption of a constant $\kappa_{250 \mu m}$ along the line of sight may not be good. 

\subsubsection{Using the known dust temperature and density profiles of L1544}

\begin{figure}
\resizebox{\hsize}{!}{\includegraphics{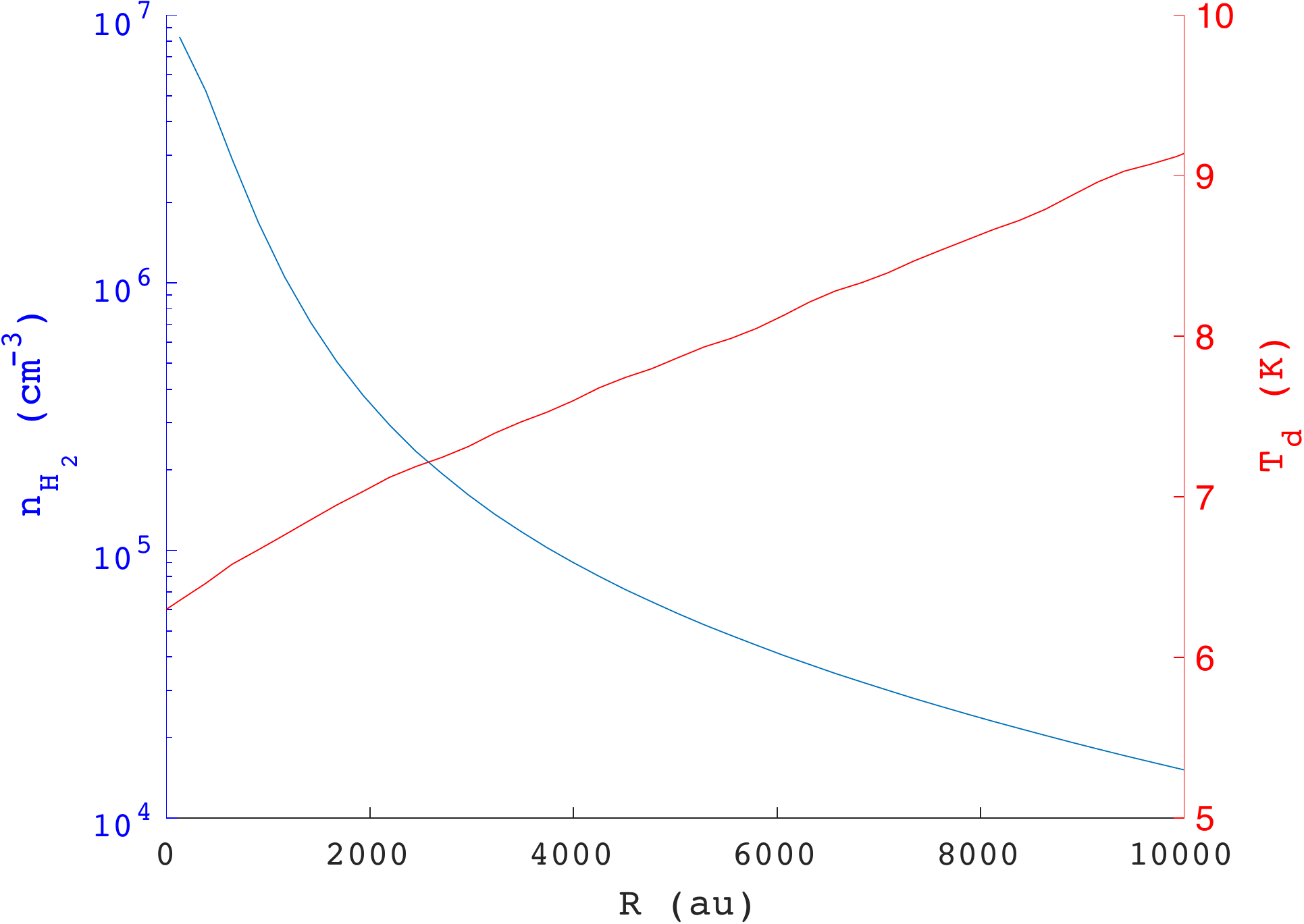}} 
\caption{Temperature (red) and density (blue) profiles of L1544 from \citetads{2015MNRAS.446.3731K}.}
\label{model}
\end{figure}

For L1544 we can make use of the radial density profile, $n_{H_{2}}(r)$, and temperature profile, $T(r)$, derived by \citetads{2015MNRAS.446.3731K}. These profiles correspond to a spherical model that follows the evolution of an unstable Bonnor-Ebert sphere, reproducing previous observations of L1544. The volume density, $n_{H_{2}}$, is related to the column density, $N_{H_2}$, as $N_{H_2} = \int_{s_{los}}^{}{n_{H_{2}} ds}$, i.e., the column density is the volume density integrated over the line of sight. $T(r)$ and $n_{H_{2}}(r)$ are shown in Fig. \ref{model}. 

The knowledge of the temperature and density of the core leaves Eq. \eqref{eq1} with only two unknown parameters: the dust opacity and spectral index. Therefore, taking into account the variation of the density and the temperature along the line of sight, and the units we use (MJy/sr), Eq. \eqref{eq1} for each pixel transforms to:

\begin{equation}
S_{\nu}(r_{ij}) = \kappa_{\nu}(r_{ij}) \mu_{H_2} m_{H}   \int_{s_{los}}^{}{B_{\nu}[T_{d}(r_{ij})]n_{H_{2}}(r_{ij}) ds},
\label{eqSS}
\end{equation} 
where $s$ is the path along the line of sight, which has a direct relation with the radius $r$, and $i$ and $j$ represent the pixel coordinates on the map.  

As for the previous section, the ratio between the surface brightnesses at 1.2 and 2 mm gives us an estimate of $\beta$, and consequently, the value of $\kappa_{\nu}$, although in this case we will derive it using:
\begin{equation}
\kappa_{\nu}(r_{ij}) = \frac{S_{\nu}(r_{ij})}{ \mu_{H_2} m_{H}   \int_{s_{los}}^{}{B_{\nu}[T_{d}(r_{ij})]n_{H_{2}(r_{ij})} ds} }.
\label{opacity_eq}
\end{equation} 
\\

This approach is more accurate than the method used in Section 3.2.1 because it does not assume a constant value for the opacity at a reference frequency, but it also has the clear wrong assumption that the spectral index and the opacity do not change \textit{along the line of sight}. This may not be the case, as if changes from the outskirts to the center of the core are present, the same change should apply along the line of sight toward the center. This approach also assumes that the spectral index does not depend on the temperature, but as we have seen in Section 1, no significant temperature dependence is expected within the temperature range relevant for L1544 \citep[e.g.]{1996ApJ...462.1026A, 1998ApJ...496.1058M,2005ApJ...633..272B,2011A&A...535A.124C,2013lcdu.confE..44D}.

\begin{figure} 
\resizebox{\hsize}{!}{\includegraphics{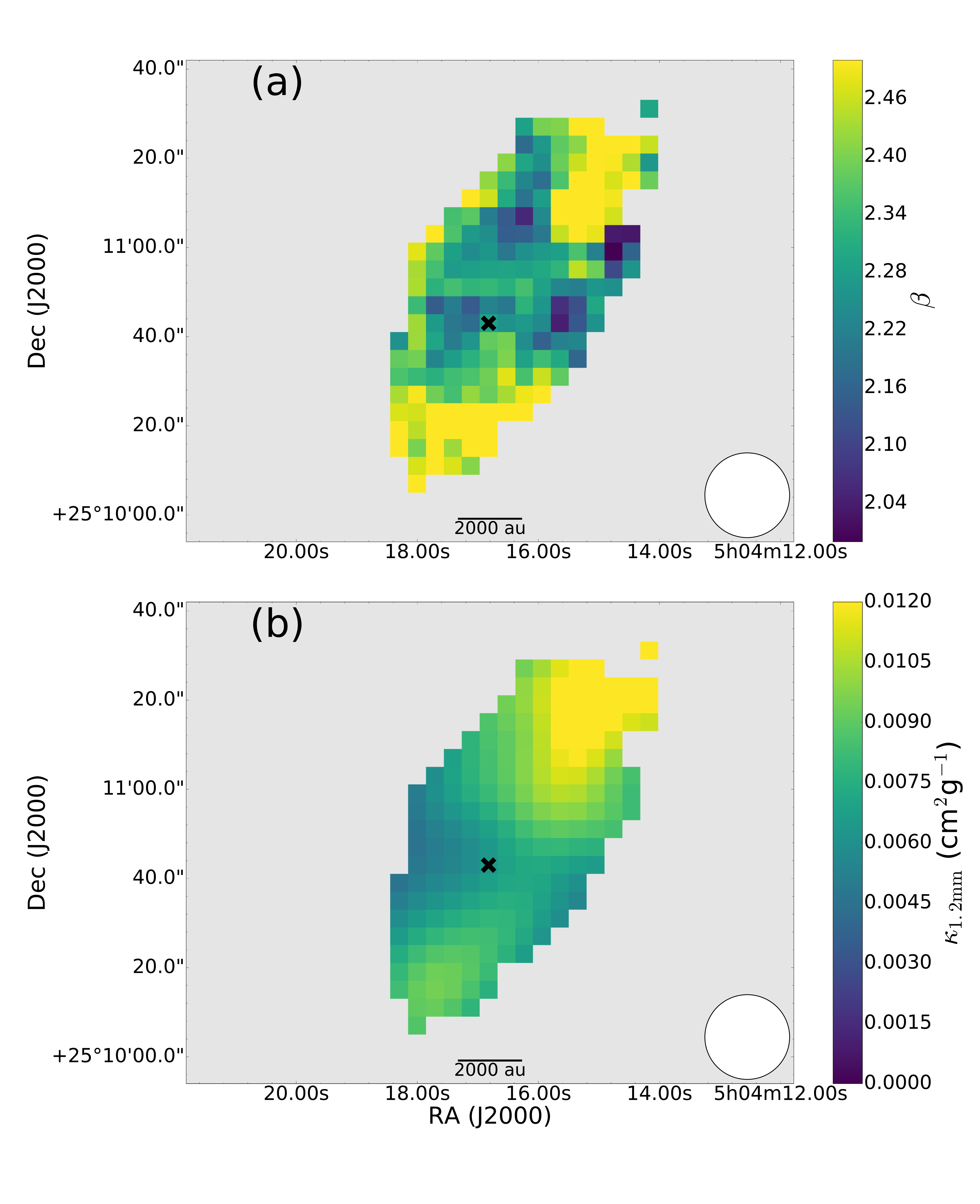}} 
\caption{Maps of the (a) spectral index and (b) the opacity at 1.2 mm derived with the method described in Section 3.2.2. The angular resolution is 19\arcsec (HPBW shown by the white circle in the bottom right corner). The crosses mark the peak of the emission corresponding to the final 1.2 mm map convolved to 19\arcsec.}
\label{kappa-beta-model}
\end{figure}

The $\beta$ and $\kappa_{1.2mm}$ maps are shown in Fig. \ref{kappa-beta-model}. The behaviour of the spectral index seems not to differ much from the one presented in Section 3.2.1, where a constant temperature was assumed. However, here the spatial variations are not significant, as they are negligible within the errors ($\sim$25\%). This is due to the fact that the ratio of the term $\int_{s_{los}}^{}{B_{\nu}[T_{d}(r_{ij})]n_{H_{2}}(r) ds}$ at the two different frequencies does not differ significantly from the one computed assuming a constant temperature of 10 K. On the other hand, the opacity value changes significantly, $\sim$60\% with uncertainties of 20\%. Thus, the overall $\kappa_{1.2mm}$ value dispersion is two times larger than before, and surprisingly, the opacity is now showing a decreasing trend towards the central regions, contrary to the expectation of grain growth. 

The simplest explanation of this apparently contradictory result is that the model assumes spherical symmetry, while L1544 is clearly an elongated core, with an aspect ratio of about 2. This implies that the model cannot be used for the outer parts of the cloud, in particular along the major axis. Moreover, the derivation of the spectral index using the method of the fluxes ratio with only two wavelengths in the Rayleigh-Jeans limit leads to inaccurate values due to the uncertainties on the fluxes \citepads{2009ApJ...696.2234S}. 

\subsection{Spectral index and opacity of the dust using NIKA and SPIRE}

\begin{figure} 
\begin{center}
\includegraphics[scale=0.5]{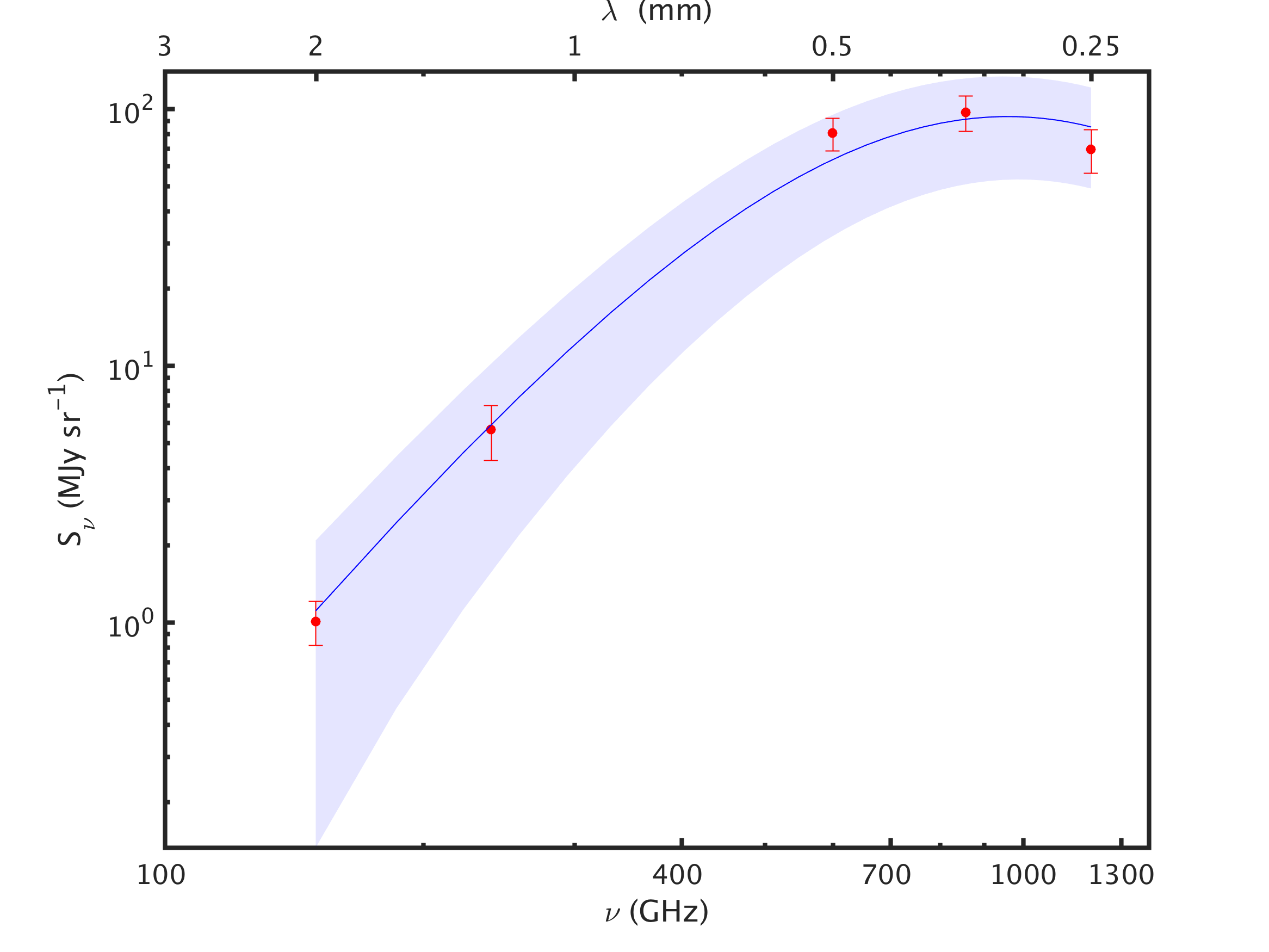}
\end{center}
\caption{Fit of the SED to the 5 spectral windows available at 250, 350, 500, 1250 and 2000 $\mu$m at the peak emission when convolving and regridding all windows to the resolution of the 500 $\mu$m band. The error bars indicate the weights used in the fitting, that correspond to the uncertainties and noise associated to the data. The shadowed blue region show the 95 \% confidence intervals of the fitted parameters.}
\label{SED_fit}
\end{figure}

Due to the complex and unknown dependence of the spectral index with the temperature, the uncertainties in the surface brightnesses, and the spherically symmetric model, here we consider a different approach. The modeled densities and temperature profiles are now used to predict the emission at 1.2 and 2 mm, fixing the values of the opacity and the spectral index of the dust.
The adopted values for the opacity and the spectral index are derived from the best least squares fit to the surface brightnesses in the five spectral windows available (3 from Herschel/SPIRE and 2 from NIKA) toward the peak emission of L1544 at each band, assuming that the opacity and the spectral index are independent, as possible degeneracies between both parameters are beyond the scope of this paper and not well-known. Before fitting, the five maps have been convolved and regridded to the lowest resolution, i.e. the 500 $\mu$m band ($\sim$38.5\arcsec beam and 14\arcsec pixel size), and the Herschel/SPIRE bands have been color corrected assuming an extended source with a spectral index of 2.0. From the model, we derive the term $\int_{s_{los}}^{}{B_{\nu}[T_{d}(r_{ij})]n_{H_{2}}(r) ds}$  for each pixel in the maps at each frequency and smooth this to the common resolution to introduce in the fitting the corresponding terms which depend on temperature and density in the center. A least squares fit is performed to find which values of the opacity and spectral index better match observations. This procedure is iterative: the spectral index and the corresponding color correction of the SPIRE bands are varied until a convergence is found. The resultant spectral energy distribution (SED) is shown in Fig. \ref{SED_fit}. The values of $\kappa_{250 \mu m}$ and $\beta$ found with this procedure are: \\

$\kappa_{250\mu m} = 0.2 \pm 0.1$ cm$^2$g$^{-1}$ and $\beta=2.3 \pm 0.4$,
\\
\\
where the errors indicate the 95\% confidence intervals of the estimates of the fit. Within the uncertainties, this value for the opacity is in agreement with those calculated by \citetads{1994A&A...291..943O} for thin and thick ice mantles. The spectral index is however significantly larger than the one derived by Planck for L1544: 1.608$\pm$0.007 \citepads{2016A&A...594A..10P}. This difference could be due to the significantly larger beam for the high-resolution (7.5 arcmin) thermal dust emission map derived using Planck frequencies above 143 GHz. In the derivation of the spectral index from Planck, a constant dust temperature of 16.55$\pm$0.08 K in a region of 1$^{\circ}\times$1$^{\circ}$ of size is assumed. This temperature is higher than that observed with ammonia by \citetads{2007A&A...470..221C}. The sensitivity of Planck to large emission and its resolution imply that the variations on the temperature are averaged over a bigger structure along the line of sight, dominated by the envelope,  increasing the observed temperature and tracing better the outer part of the core. Moreover, if no temperature variation along the line of sight is taken into account, $\beta$ is expected to be underestimated \citepads{2015A&A...584A..94J}, therefore we adopt our derived value of beta as representative for L1544. 

\begin{figure} 
\begin{center}
\includegraphics[scale=0.45]{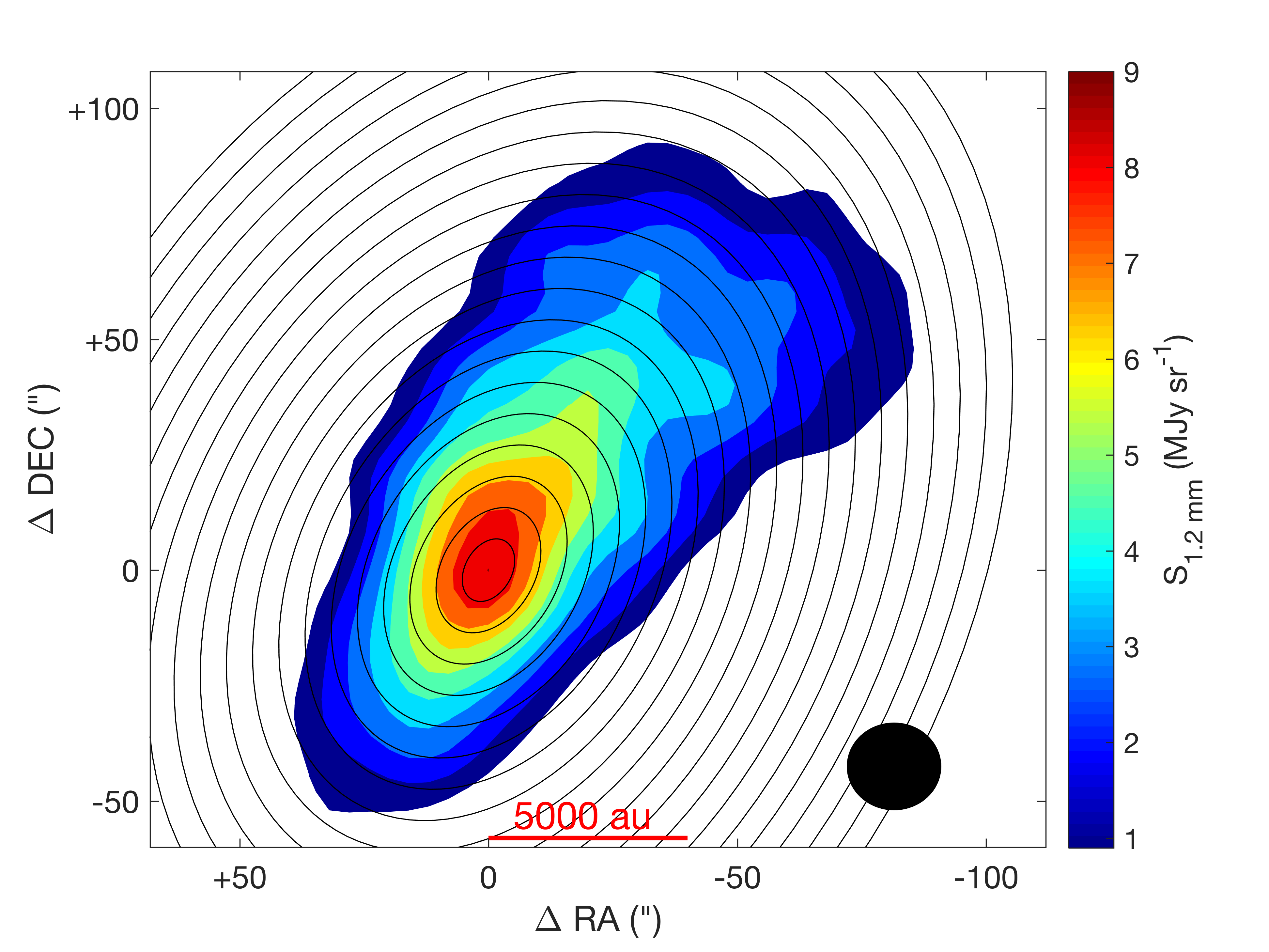}
\end{center}
\caption{Ellipses where the emission has been averaged for the 1.2 mm and 2 mm maps on top of the 1.2 mm continuum map of L1544. Only the colored region has been included in the average. The resolution is 19\arcsec (HPBW shown by the black circle in the bottom right corner). For Herschel/SPIRE the distance between concentric ellipses is larger due to a larger pixel size and resolution. For both NIKA and SPIRE, the distance between ellipses is chosen to be $\sim$1.8 pixels, averaging in a 1 pixel wide ring.}
\label{elip}
\end{figure}

\begin{figure} 
\begin{center}
\includegraphics[scale=0.43]{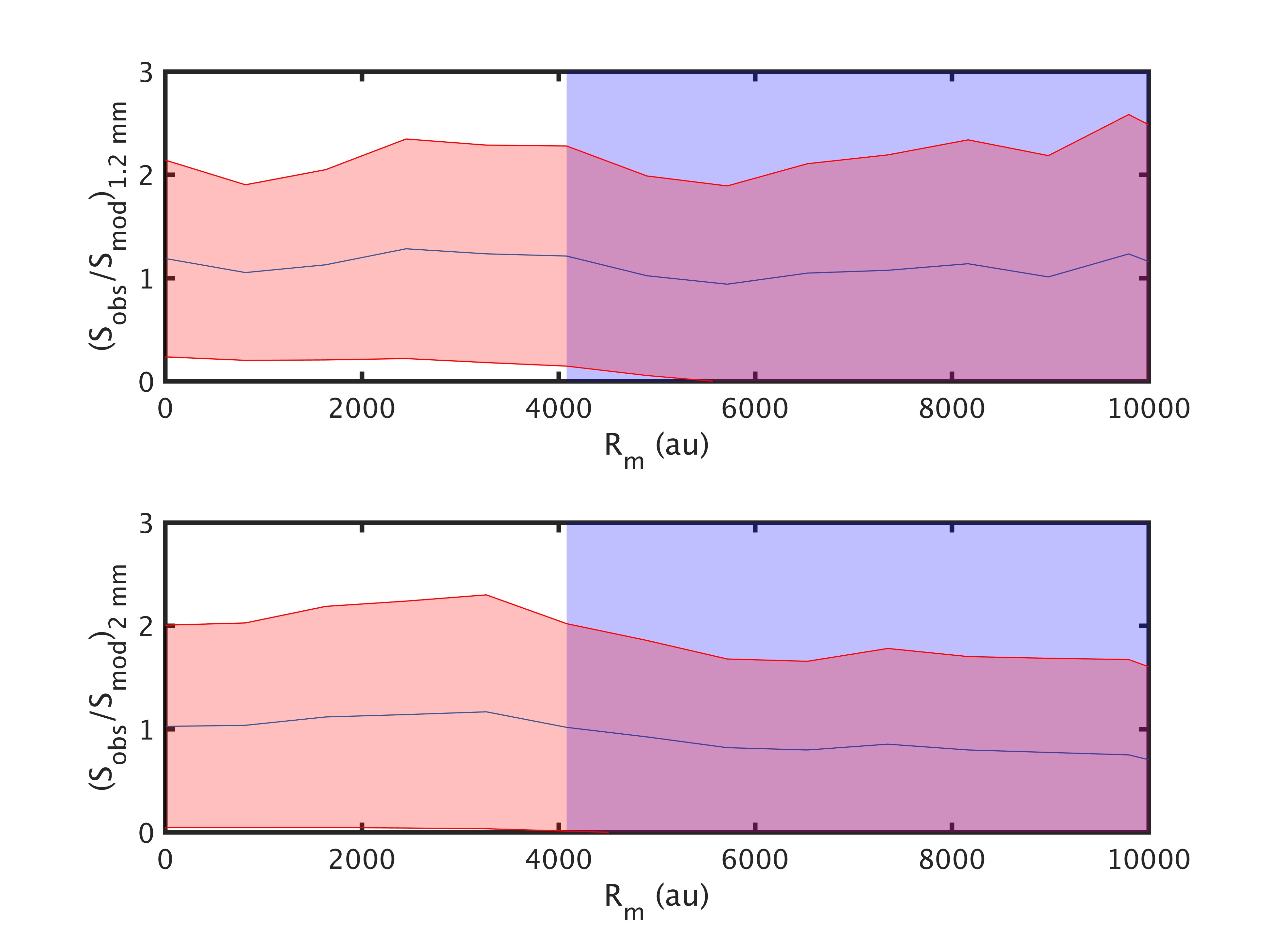}
\end{center}
\caption{Ratio between the observed and the modeled surface brightnesses, at 1.2 mm (top) and 2 mm (bottom), as a function of geometric mean radius of the ellipses in Fig. \ref{elip}, using the temperature and density profiles from \citetads{2015MNRAS.446.3731K} and an opacity value of $\kappa_{250\mu m} = 0.2 \pm 0.1$ cm$^2$g$^{-1}$ and $\beta=2.3 \pm 0.4$. The shadowed red region represents the uncertainties in the fluxes, the opacity and the spectral index. The shaded blue region indicates the zone where the observed surface brightness is below 3 $\sigma$.}
\label{profiles_mm}
\end{figure}

\begin{figure} 
\begin{center}
\includegraphics[scale=0.43]{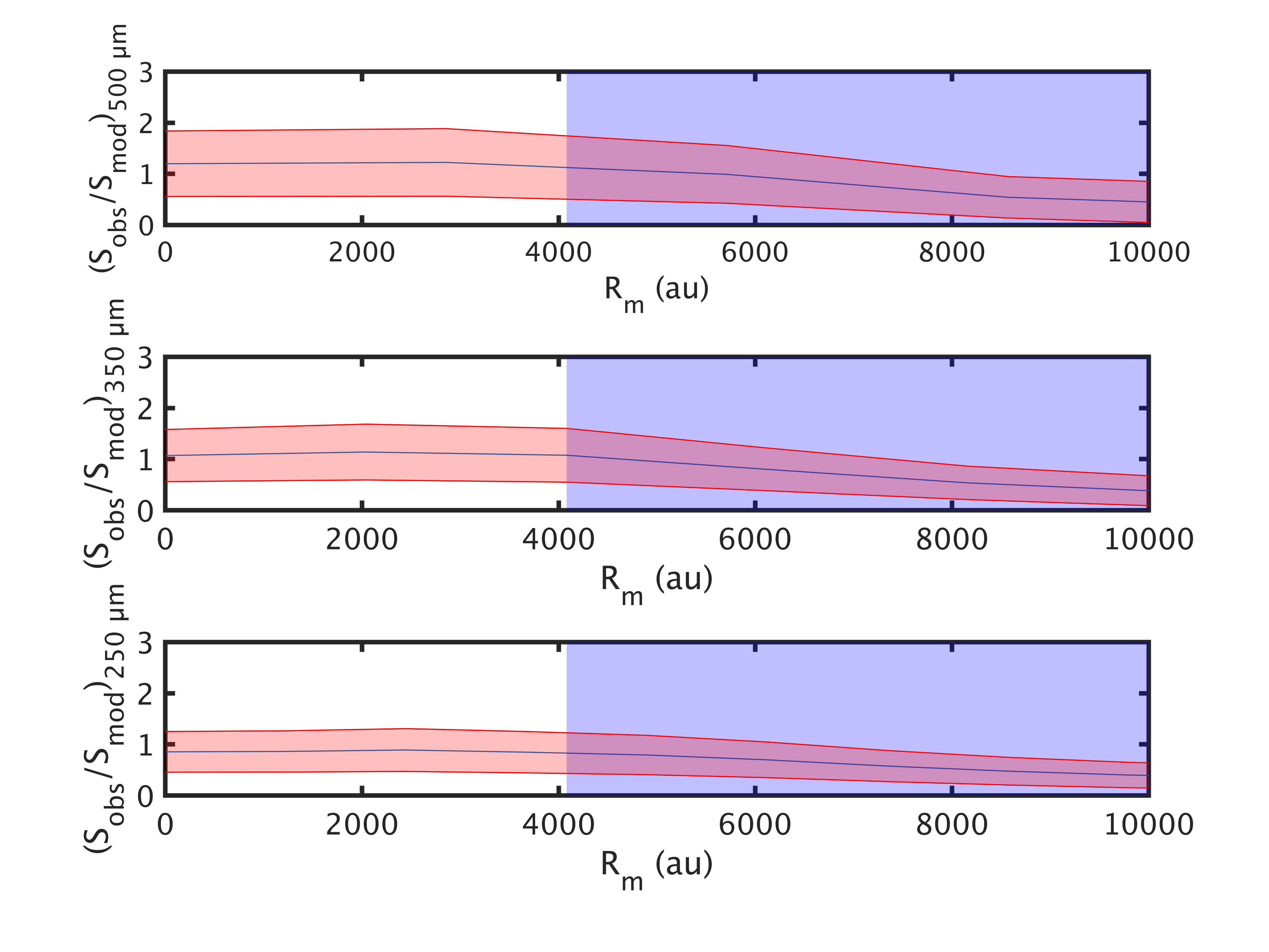}
\end{center}
\caption{Ratio between the observed and the modeled surface brightnesses, at 500 $\mu$m (top), 350 $\mu$m (middle) and 250 $\mu$m (bottom), as a function of the geometric mean radius of the ellipses as explained in Fig. \ref{elip}, using the temperature and density profiles from \citetads{2015MNRAS.446.3731K} and an opacity value of $\kappa_{250\mu m} = 0.2 \pm 0.1$ cm$^2$g$^{-1}$ and $\beta=2.3 \pm 0.4$. The shadowed red region represents the uncertainties in the fluxes, the opacity and the spectral index. At 250 $\mu$m, the dashed lines coincide with the solid blue line, when $\beta=2.3$, as the reference wavelength for the opacity is at 250 $\mu$m. The shaded blue region indicates the zone where the millimeter observed surface brightness is below 3 $\sigma$, as the artificial filtering of Herschel/SPIRE relies on the shape of the 1.2 mm emission (see Section 2.2 and Appendix A).}
\label{profiles_spire}
\end{figure}

To make the comparison between the modeled emission, which is for a sphere, and the observed one, we average the emission of the core in ellipses at different distances from the center with an inclination of $\sim$65$^{\circ}$ (see Fig. \ref{elip}). Only emission within the colored map in Fig. \ref{elip}, which corresponds to positive surface brightnesses, has been considered. The results are shown in Fig. \ref{profiles_mm} for NIKA and in Fig. \ref{profiles_spire} for SPIRE, where the ratio between the observed and modeled surface brightnesses is reported as a function of radius (for the observed emission this radius corresponds to the geometric mean of the major and minor axes of the ellipses). The resolution is the native resolution of each band, which corresponds to 18.5\arcsec, 12.5\arcsec, 38.5\arcsec, 26.8\arcsec and 20.3\arcsec, at 2 mm, 1.2 mm, 500 $\mu$m, 350 $\mu$m and 250 $\mu$m, respectively. The agreement is within a factor of 2 in all bands, which leads to the conclusion that there is no need to modify the opacity or the spectral index towards the central regions to reproduce the observed surface brightnesses. We have also checked the dependence of this ratio with the value of the spectral index and the opacity: the mm-wavelengths are sensitive to the variation of $\beta$ and $\kappa_{250 \mu m}$, increasing and decreasing the surface brightness by a factor or 2 or more. However, if the spectral index and the opacity increase at the same time within the 95\% confidence intervals, the ratio between the observed and the estimated surface brightness maintains constant around 1 within the errors. This is because an increase of the spectral index decreases the expected surface brightness at mm-wavelength, but an increase of the opacity can compensate this effect and increases the prediction by the same amount. Therefore, variations of these two parameters within the uncertainties (50\% and 20\% for the opacity and the spectral index, respectively) cannot be detected with our data. The decrease in the ratio between the observed and modeled surface brightness at the outskirts of the core in the case of SPIRE is due to the harsh filtering applied to the data, and the error associated with this procedure is large at R > 4000 au because the filtering relies on the shape of the 1.2 mm emission, which goes below the 3 $\sigma$ level at R > 4000 au (see Section 2.2 and Appendix A). 

These results indicate that the emission is consistent with a constant opacity, at the resolution and sensitivity achieved with NIKA. This is in apparent contradiction with \citetads{2010MNRAS.402.1625K}, who needed to increase the opacity towards the central regions to reproduce the drop in the temperature measured by \citetads{2007A&A...470..221C} using interferometric observations. As explained in Section 4.3, higher angular resolution observations are needed to test the prediction made by \citetads{2010MNRAS.402.1625K}. 

\section{Model predictions on grain growth and comparison with our data} \label{model_sec}

The theoretical studies on grain evolution of \citetads{2009A&A...502..845O, 2011A&A...532A..43O} show that little grain growth takes place when the cloud lifetime is similar to the free-fall time, because the process of coagulation takes longer. However, pre-stellar cores may span a range of lifetimes which significantly exceeds the free-fall time \citep[see e.g.]{2014Natur.516..219B, 2015ApJ...804...98K, 2015A&A...584A..91K}. If there are processes sustaining the cloud against gravitational collapse (e.g. turbulence and/or magnetic fields), some dust coagulation could take place. \citetads{2009A&A...502..845O,2011A&A...532A..43O} show that about 100 $\mu$m dust grains can form if gas densities of 10$^5$ cm$^{-3}$ can be maintained for 10 Myr, while after 1 Myr dust grains can grow up to $\sim$4 $\mu$m. 

In the following, we present an analytical solution \citepads{2004ASPC..309..369B} to estimate grain growth. \citetads{2009A&A...502..845O} compared this analytical solution with their Monte Carlo simulation, finding a very good match. We apply this model to the pre-stellar core L1544. As mentioned, L1544 is a very well-known pre-stellar core, centrally concentrated and with high central densities ($\sim$10$^6$ cm$^{-3}$ within a radius of 500 au, see Fig. \ref{model}) which makes it unique and ideal for this study. We consider two different cases: a static core, with the physical structure given in Fig. \ref{model}, and a dynamical core, based on the quasi-equilibrium contraction model of \citetads{2015MNRAS.446.3731K} which best reproduce spectroscopic data.

\subsection{A simple analytical model}

We have used a simple analytical model derived by \citetads{2004ASPC..309..369B}, and checked by \citetads{2009A&A...502..845O}, who compared it with their Monte Carlo simulation and demonstrated it is valid for average molecular cloud conditions, with densities of 10$^5$ cm$^{-3}$, temperature of 10 K, and at intermediate velocity regimes\footnote{The "intermediate velocity regime" is defined as the velocity range within which the gas-dust friction time is larger than the eddy turnover time.  In these conditions, the velocity scales with the square root of the largest particle friction time \citepads{2009A&A...502..845O, 2007A&A...466..413O}.}  \citepads{1993A&A...280..617O, 2009A&A...502..845O, 2007A&A...466..413O}. The analytical model is based on the assumption of a monodisperse distribution of spherical particles. As the particles coagulate, they will form bigger grains made out of N monomers, with a characteristic filling factor, which describes their shape (fractal or compact). Here we use the terminology used by \citetads{2009A&A...502..845O}, as well as their parameters and equations derived from their simulations. 

The initial evolution of the process of grain growth through grain-grain collisions can be divided into two different regimes: the fractal regime and the compaction regime \citepads{2009A&A...502..845O,2004ASPC..309..369B}. The fractal regime is characterized by growth with no visible restructuring of the grain, also it includes the so-called hit-and-stick regime. By contrast, during the compaction regime dust grains suffer changes in their structure. The transition point from one regime to another is determined by the parameter $N_1$ \citep[see Appendix A from]{2009A&A...502..845O}, which strongly depends on the grain size. The bigger the grain, the lower $N_1$ and the easier to reach the compact limit. $N_2$ \citep[see Appendix A from]{2009A&A...502..845O} instead is going to define the transition point between the compaction regime and the fragmentation regime, where grains start to lose monomers in grain-grain collision. When this regime is reached, it is assumed that the grains do not grow more and the growth is compensated with fragmentation.
When grains are close to this regime in high density environments, the simple model overestimates the growth compared to \citetads{2009A&A...502..845O}, due to the high dependency of $N_1$ with density.

To define the transition points $N_1$ and $N_2$, we apply the same criteria as \citetads{2009A&A...502..845O}. In their Appendix A one can find the complete expressions for $N_1$ and $N_2$, which follow: 
\begin{equation}
N_1 \propto \left(\frac{n}{10^5 \, \mathrm{cm^{-3}}}\right)^{3.75}\left(\frac{a_0}{0.1 \, \mathrm{\mu m}}\right)^{-22.5} ,
\end{equation}
and
\begin{equation}
\frac{N_2}{N_1} \propto \left(\frac{a_0}{0.1 \, \mathrm{\mu m}}\right) .
\end{equation}

Please note that $n$ here is the number density of gas molecules, related to $n_{H_2}$ as $n \simeq n_H/1.7 = 2 n_{H_2}/1.7$, and $a_0$ is the initial grain size.

As derived in \citetads{2004ASPC..309..369B} and used by \citetads{2009A&A...502..845O}, the equation to solve is:

\begin{equation}
\frac{\mathrm{d}N}{\mathrm{d}t} =  \frac{N}{t_{coll}} = \frac{N^{5/6} \phi_{\sigma}^{-1/3}}{t_{coll,0}} ,
\end{equation}
where $N$ is the number of monomers and $\phi_{\sigma}$ is the filling factor of the dust grain, a quantity that indicates the fluffiness of the grains. Its value is assumed to be $N^{-3/10}$ for the fractal regime, and $N_1^{-3/10}$ for the compact regime. For the fragmentation regime we adopted the value 0.33 \citepads{2009A&A...502..845O, 2004PhRvL..93k5503B}. In our calculations the fragmentation regime is reached in the center of the core if we maintain the observed central density of L1544 constant with time, for $t\gtrsim0.7$ Myrs. However, as shown in section 4.3, this regime is not reached when a more realistic dynamical evolution is taken into account.

\subsection{Deriving optical properties}

Once the final expressions for the number of monomers that can coagulate after a certain time of evolution are derived, the corresponding grain size and the grain opacity can be calculated. 

To calculate dust grain opacities, we use the code described in \citetads{2016A&A...586A.103W}\footnote{The code is available at \url{https://dianaproject.wp.st-andrews.ac.uk/}}. More information can be found as well in \citetads{1981ApOpt..20.3657T, 2005Icar..179..158M, 1995A&A...300..503D} and \citetads{1996MNRAS.282.1321Z}. We do not enter into the details of dust grain composition and ice mantles in the calculation of the dust opacity, as only differential values are of importance in our study and the option of adding ice mantles to dust grains has not been yet implemented in the online version of the code presented by \citetads{2016A&A...586A.103W}. However, ice mantles should be consider in further detailed analysis. We will refer to this code as the opacity calculator.

For relating the number of monomers with the final grain size, we follow the suggestion of \citetads{2011A&A...532A..43O} and consider the projected surface equivalent radius as the final particle size introduced into the opacity calculator:

\begin{equation}
a_{\sigma} = \left( a_0 ^3 \frac{N}{\phi_{\sigma}} \right) ^{1/3} .
\end{equation}

This is calculated for every value of $N$ obtained. 

Finally, the size distribution is approximated by a power-law function of the form $n(a)da \propto a^{-p}$ \citepads{2016A&A...586A.103W} and then integrated between the maximum and minimum grain size present along the line of sight using the opacity calculator. Therefore, we give as input in the opacity calculator the different grain sizes distributions.

\subsection{Results}

Grain growth depends on density, temperature and dust grain properties. As the temperature and density profiles of L1544 are relatively well known, we can estimate the grain growth in different points of the cloud and predict how much variation of grain sizes is present across the core. We then convolve the modeled distribution with the instrumental beam size to simulate observations.  

We consider two scenarios: one where the cloud has not evolved and has maintained its density structure over time, hereafter static cloud, and another one where the cloud has evolved and changed its structure with time, hereafter dynamic cloud. In the dynamic scenario, we use the density profiles at different times of evolution of a Bonnor-Ebert sphere in quasi-equilibrium contraction, starting from an unstable dynamical equilibrium state with an initial density of 10$^4$ cm$^{-3}$, as deduced by \citetads{2015MNRAS.446.3731K} (their Fig. 3), and stopping when the observed L1544 central density is reached (\textit{t}$\sim$1.06 Myrs). The density change is followed in steps, by interpolating the various density profiles at different times. Integrating over an interval of time, we can then obtain the grain size distribution evolution starting from an assumed monodisperse initial distribution, and the opacity corresponding to that distribution. 

In both cases, the monodisperse distribution starts with $a_0 = 0.1$ $\mu$m, and will follow the grain evolution at every period of time in the \citetads{2015MNRAS.446.3731K} dynamical model.
The corresponding opacity maps are convolved with a beam of 20\arcsec for simulating NIKA observations and of 2\arcsec to simulate observations with the Atacama Large Millimeter and submillimeter Array (ALMA). 

In the static cloud, with the same density and temperature profiles as L1544, dust grains grow in the center up to cm-sizes in 1.06 Myrs. We consider this result not realistic for two reasons.  Firstly, it implies a constant central density of 10$^7$ cm$^{-3}$ for 1.06 Myrs, while \citetads{2010MNRAS.402.1625K} and \citetads{2015MNRAS.446.3731K} have shown that observed line profiles toward this pre-stellar core are consistent with quasi-equilibrium contraction motions; this dynamical evolution implies central densities larger than 10$^6$ cm$^{-3}$ only in the last $\sim$0.06 Myrs. Secondly, the analytical model overestimates grain growth for densities significantly higher than 10$^5$ cm$^{-3}$. For densities of 10$^6$ cm$^{-3}$ and 10$^7$ cm$^{-3}$, the analytic formula overestimates the grain growth by a factor of $\sim$3 and $\sim$50, respectively, after 1 Myr. In the more realistic dynamical cloud model, the grains reach sizes up to $\sim$230 $\mu$m in the core center after a time evolution of 1.06 Myrs. 

\begin{figure} 
\begin{center}
\includegraphics[scale=0.45]{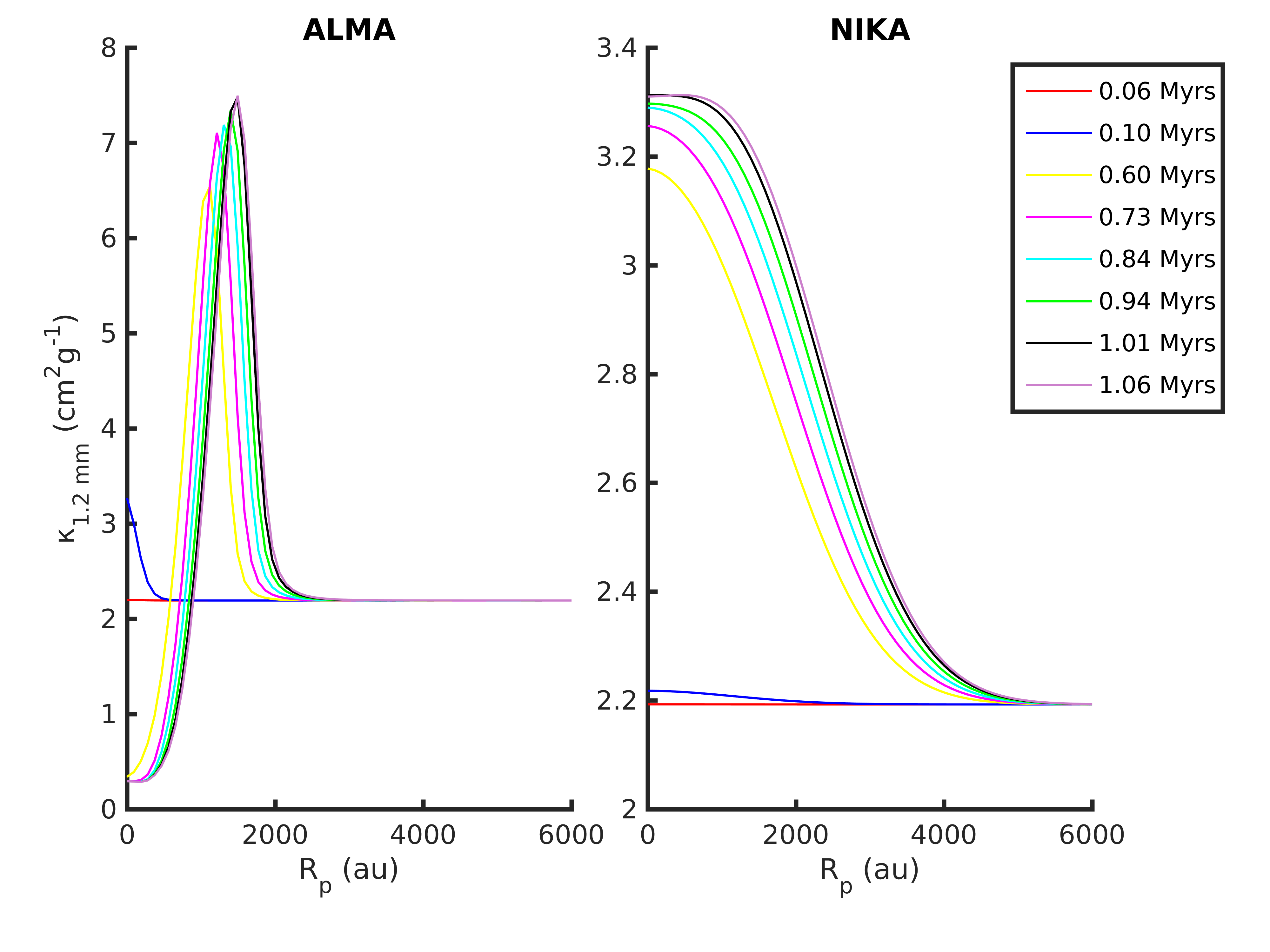}
\end{center}
\caption{Variation of the opacity with time and projected radius deduced with ALMA (on the left) and NIKA (on the right) simulated observations for the case of a static cloud. The opacity for the NIKA simulated observations changed by a factor of $\sim$1.5, while for ALMA the change corresponds to a factor of $\sim$14.}
\label{op_static}
\end{figure}

\begin{figure} 
\begin{center}
\includegraphics[scale=0.45]{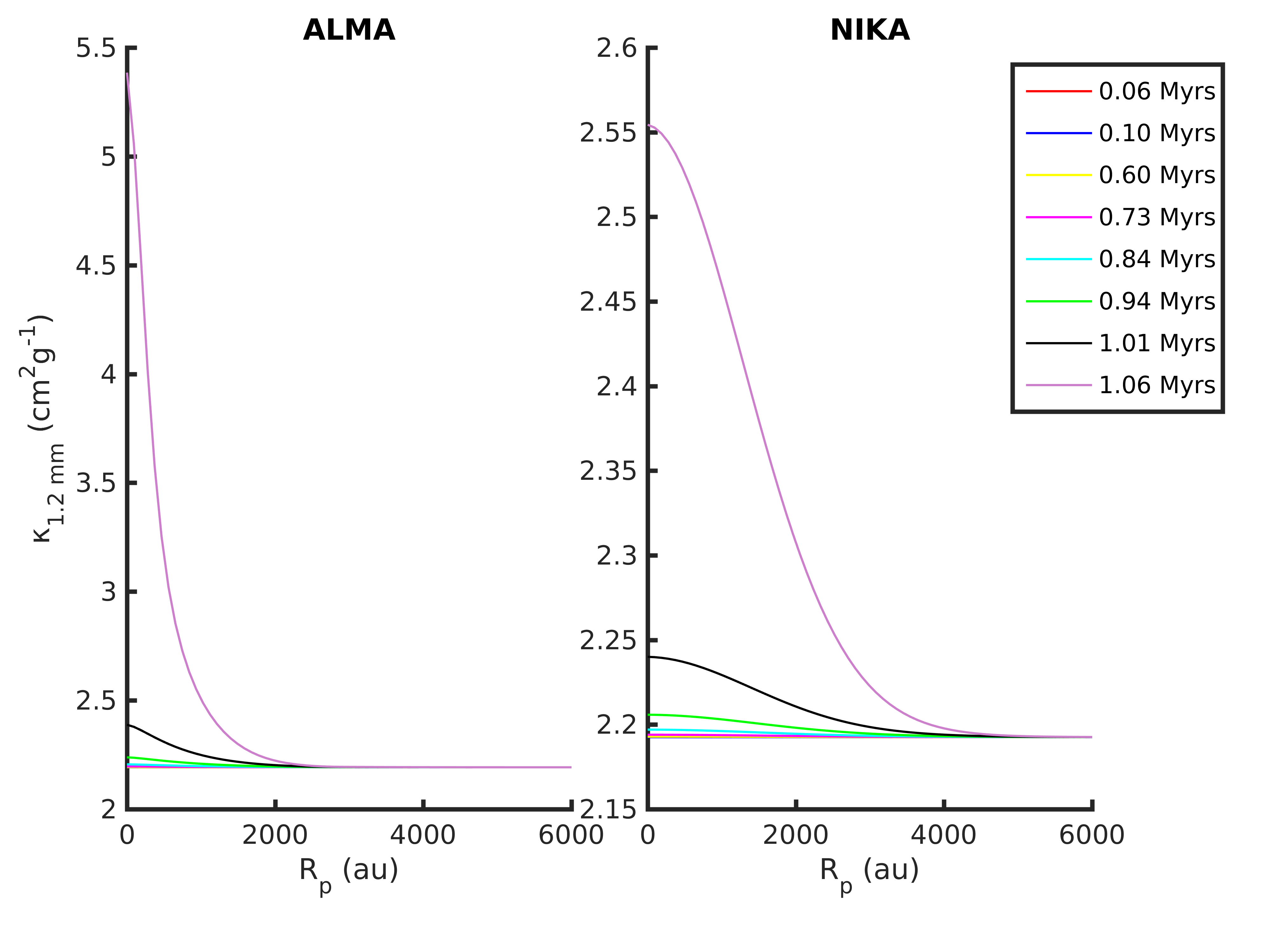}
\end{center}
\caption{Variation of the opacity with time and projected radius deduced with ALMA (on the left) and NIKA (on the right) simulated observations for the case of a dynamical cloud. The change on the opacity between the center and the outer edge for the case of NIKA is less than 15\%, while for ALMA it is a factor of $\sim$2.5 after 1.06 Myrs of evolution. }
\label{op_dynamic}
\end{figure}

The corresponding opacities at 1.2 mm, which without filtering are proportional to the expected flux, are shown in Fig. \ref{op_static} and Fig. \ref{op_dynamic} for the static and dynamical cloud, respectively. In the case of a static cloud, the predicted variation of the opacity across the cloud would have been seen with our data, as it predicts a variation of a factor of $\sim$1.5 in the opacity value at NIKA's resolution. In Fig. \ref{op_static}, it is also evident the drop in opacity when the grains are much larger than the wavelength at which the opacity is derived at ALMA's resolution, and the enhancement where $a \sim \lambda$, between 1000 and 2000 au. For the dynamic cloud (Fig. \ref{op_dynamic}), the variation at NIKA's resolution is not observable with the sensitivity reached, as it is below 15\% and the RMS of our data corresponds to $\sim$15\% of the peak emission. This result is consistent with our observations, which show that no change on the opacity is seen or needed to reproduce NIKA data. However,  Fig. \ref{op_dynamic} shows that the dust opacity change should be detected with high-resolution telescopes, like ALMA, as an opacity increase by a factor of 2.5 is predicted toward the central regions of the pre-stellar core. This is a difficult task, as ALMA would filter the extended emission, but the scale we need to look at to find the opacity change is the central 1000-2000 au, and under the right observational configurations (an angular resolution of at least $\sim$5\arcsec for obtaining an expected change in the opacity of a factor of $\sim$2, the inclusion of the Atacama Compact Array and mosaicing), this can be achieved. Therefore, we conclude that ALMA is needed to detect grain growth within pre-stellar cores similar to L1544.

\section{Conclusions} \label{conclusions}

Our work has focused on the search of grain growth towards the central regions of a contracting pre-stellar core, at the earliest phases of the star formation process. We have analyzed the emission from L1544 at 1.2 and 2 mm using the continuum camera at the 30 m IRAM telescope, NIKA. We searched for variation across L1544 on the spectral index and the opacity of the dust, as those are sensitive to grain sizes. 

In the derivation of the dust spectral index and opacity, we demonstrate that assumptions on constant dust temperature and spherical symmetry can lead to wrong conclusions about grain growth. 

We investigated the possibility to reproduce, from the model of \citetads{2015MNRAS.446.3731K}, the flux seen with NIKA using a constant opacity and spectral index. From the SED fit toward the center of the core using Herschel/SPIRE and NIKA bands, the derived values of $\kappa_{250 \mu m}$ and $\beta$ are: $\kappa_{250\mu m} = 0.2 \pm 0.1$ cm$^2$g$^{-1}$ and $\beta=2.3 \pm 0.4$. We find that there is no need to increase the opacity towards the central regions to be able to reproduce the observed fluxes. This means that no signatures of grain growth are present towards L1544, within the limits of our data. 

Finally, a simple analytical model of grain growth applied to L1544 shows that NIKA observations cannot detect opacity changes across the core and that only interferometers, in particular ALMA, can provide information about grain growth toward the central regions of L1544. 

To know the grain size distribution in the earliest phases of star formation is an important issue, as the presence of grain growth might affect the future formation of protoplanetary disks and their physical and chemical evolution. In the future, we plan to use ALMA and the Karl G. Jansky Very Large Array (JVLA) data to test our prediction and to extend our study to other dense cores embedded in different environments.

\begin{acknowledgements}

The authors thank the anonymous referee for the careful reading and useful comments, Nicolas Billot for his support and help in the data reduction process, as well as IRAM 30m staff, and Jonathan C. Tan and Silvia Spezzano for their useful suggestions. ACT, PC, and JEP acknowledge the financial support of the European Research Council (ERC; project PALs 320620).

\end{acknowledgements}
\bibliography{biblio}
\bibliographystyle{aa}

\begin{appendix}

\section{NIKA and Herschel/SPIRE filtering}

\begin{figure} 
\begin{center}
\includegraphics[scale=0.5]{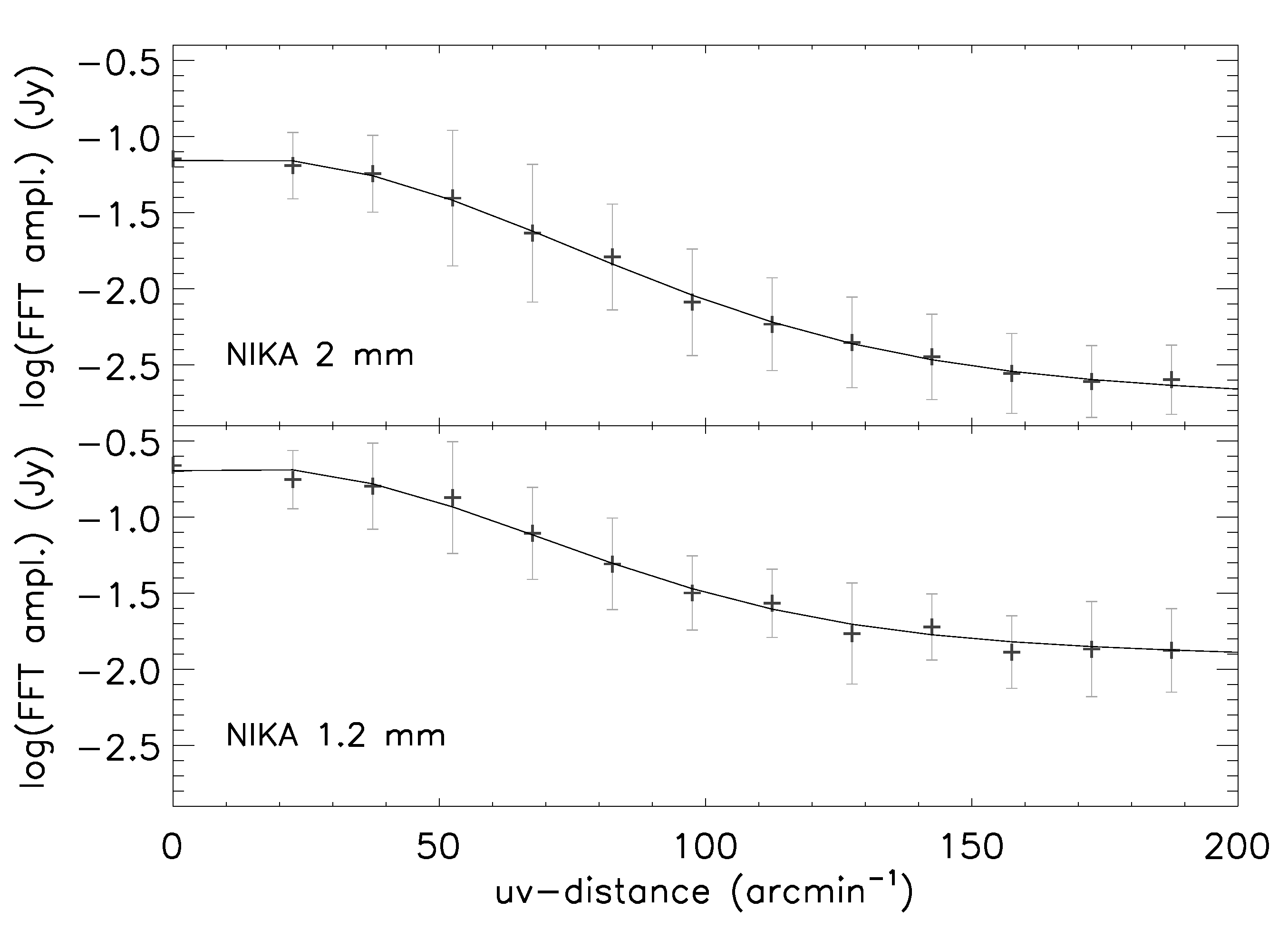}
\end{center}
\caption{Fourier amplitude profiles of the 1.2 mm (on the top) and 2 mm (on the bottom) data in the frequency domain. The profiles are obtained by averaging in rings of 1 pixel of width at different distances from the center of the Fourier transformed emission maps. The profile at 2 mm is used for deriving a mask which is applied to the Herschel/SPIRE data in order to filter out the large-scale emission.}
\label{power}
\end{figure}

\begin{figure*} 
\begin{center}
\includegraphics[scale=0.08]{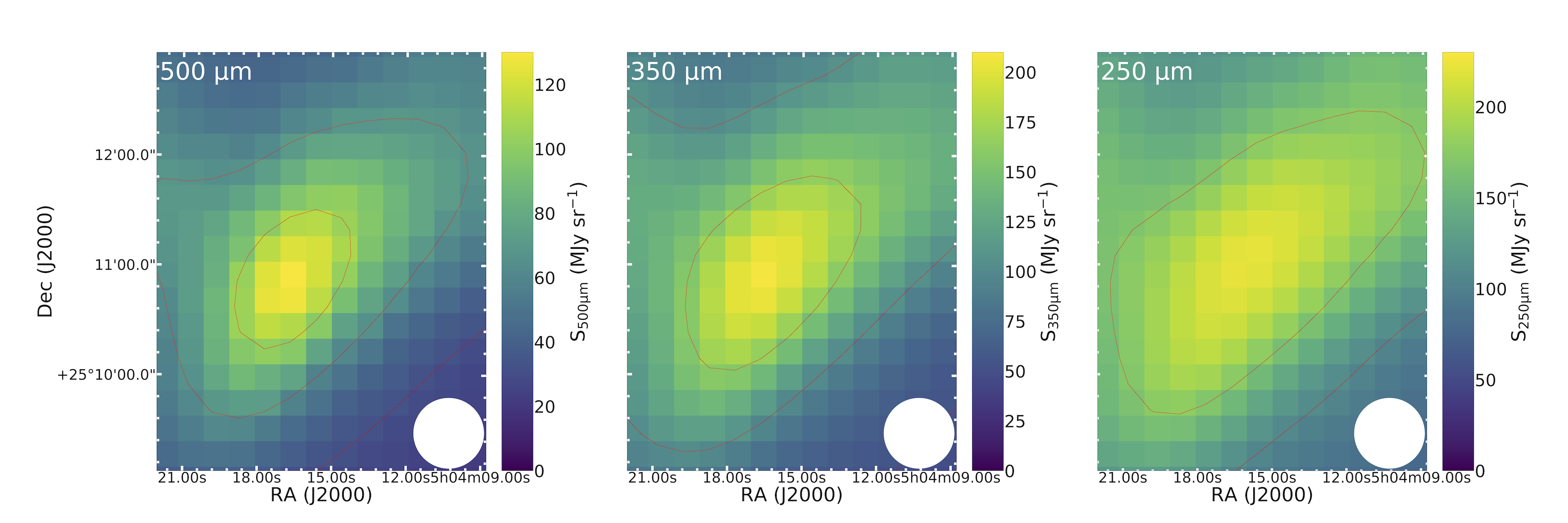}
\end{center}
\caption{The original Herschel/SPIRE maps after smoothing to the lowest resolution and before applying any filtering method. The contours represent steps of 25\% of the peak emission at each band. The HPBW, which corresponds to the larger resolution of Herschel/SPIRE (38.5\arcsec), is shown by the white circle in the bottom right corner.}
\label{maps_spire_original}
\end{figure*}

\citetads{2016A&A...588A..30S} presented a way to filter, based on the work from \citetads{2015MNRAS.450.4043W}, Herschel/SPIRE large scale emission by cutting off the low spatial frequencies in the Fourier domain based on their GISMO 2 mm data. For studying this method with our data, we smooth all the bands to the lowest resolution ($\sim$38.5\arcsec, corresponding to the 500 $\mu$m SPIRE band, these maps can be seen in Fig. \ref{maps_spire_original}) and Fourier transform all the maps. We then obtain the radial amplitude profile of the 2 mm frequency domain maps by averaging in rings one pixel wide, as done in \citetads{2016A&A...588A..30S}. This profile is shown in Fig. \ref{power}, together with the one at 1.2 mm for comparison. This profile is then fitted with an exponential function, and the fit is used to construct a mask for SPIRE maps in the same manner as described in \citetads{2016A&A...588A..30S}.

There would not be much difference between using the 1.2 mm map and the 2 mm map as a mask, as both profiles have identical slopes within the errors (see Fig. \ref{power}). This is also an indication of similar filtering in both bands, as expected from the similarity of the fields-of-view and as predicted by \citetads{2015A&A...576A..12A}. 

\begin{figure*} 
\begin{center}
\includegraphics[scale=0.08]{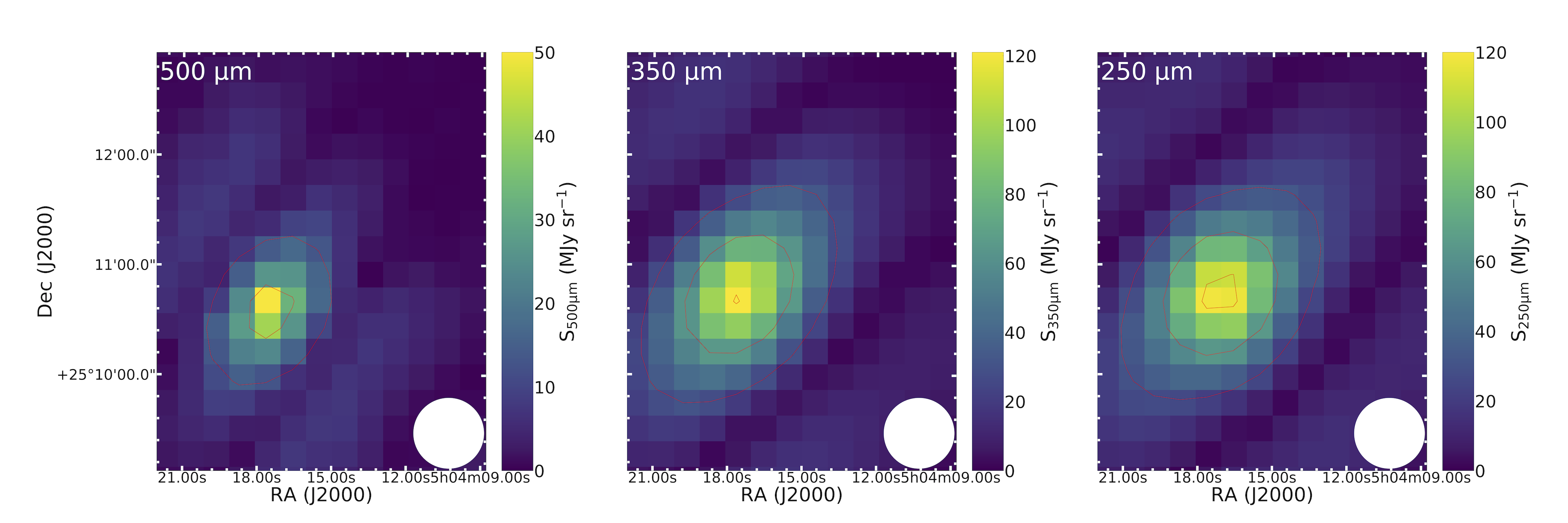}
\end{center}
\caption{The resulting Herschel/SPIRE maps after applying the filtering method from \citetads{2016A&A...588A..30S}. The contours represent steps of 25\% of the peak emission at each band. The emission is seen to be centrally concentrated in the central pixels.}
\label{maps_spire_f1}
\end{figure*}

\begin{figure*} 
\begin{center}
\includegraphics[scale=0.08]{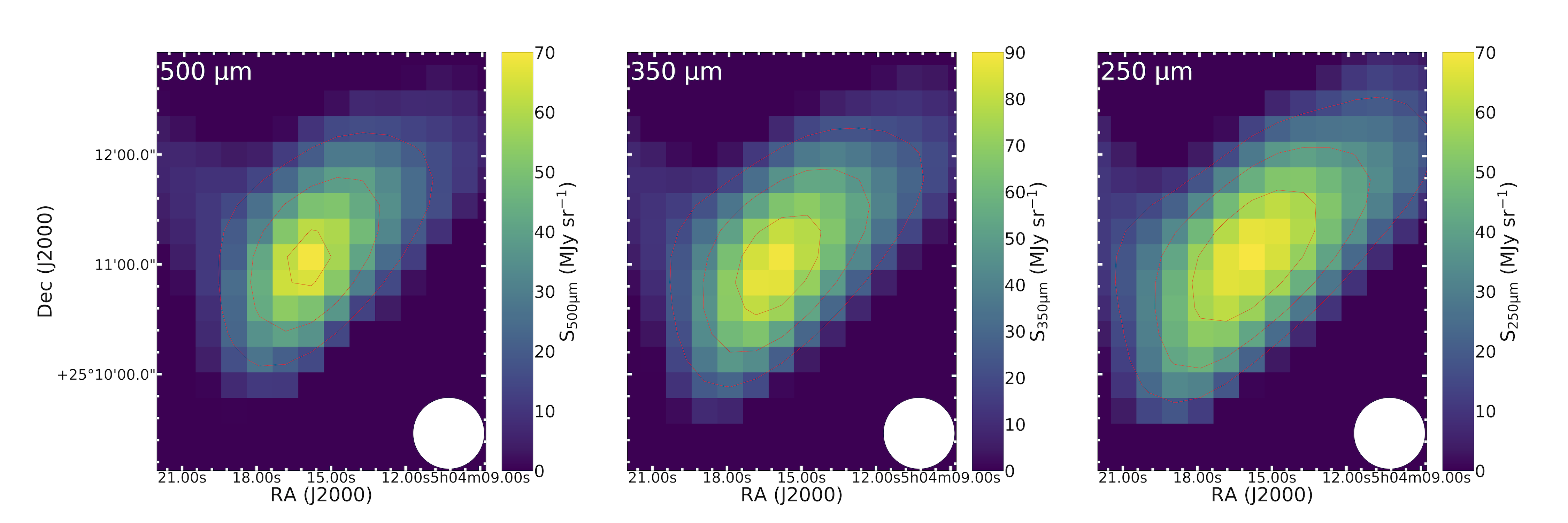}
\end{center}
\caption{The resulting Herschel/SPIRE maps after applying the filtering method used in this work, extracting the emission around the core where no emission is detected in mm-wavelengths. The contours represent steps of 25\% of the peak emission at each band.}
\label{maps_spire_f2}
\end{figure*}

The final filtered SPIRE maps are shown in Fig. \ref{maps_spire_f1} for the Fourier transform method and in Fig. \ref{maps_spire_f2} for the method used in this work (see Section 2.2), where an aperture is chosen for subtracting the emission from the original maps. In the first case, they show very centrally concentrated emission, with a decrease on the peak surface brightness of 40, 60 and 50\% for the 500, 350 and 250 $\mu$m bands, respectively, with respect to the original SPIRE maps. The peak emission we obtained with the aperture filtering method, compared with the Fourier transform method results, gives differences of 30, 25, and 40\% approximately for the 500, 350 and 250 $\mu$m bands, respectively. As the two shorter wavelengths bands see an increase in surface brightness, there is an increase in the spectral index and the opacity, obtaining 2.5 $\pm$ 0.6 and 0.28 $\pm$ 0.14 cm$^2$g$^{-1}$, respectively. Although this is still within our uncertainties, this opacity and spectral index are less good in reproducing the observed NIKA millimeter emission: the ratio between the observed emission and the model stays constant around 1.3 in both cases. This difference can be due to the too large concentration of the emission in the central pixel and this may be a source of error. It could be artificially increasing the peak emission, making the slope of the spectral energy distribution increase as well. We therefore used the method that reproduces better the millimeter observations, although both methods give similar results within the uncertainties.

\end{appendix}

\end{document}